\documentclass[12pt]{revtex4}
\usepackage{citesort}
\usepackage{hyperref}
\pdfoutput=1

\begin{document}

\title{On the sound of snapping shrimp}

\author{Michel Versluis$^1$, Barbara Schmitz$^2$, Anna von der Heydt$^1$, 
and Detlef Lohse$^2$}

\affiliation{
$^1$Physics of Fluids Group, 
Department of Science and Technology, 
J.M. Burgers Center for Fluid Dynamics, and Impact-Institute, \\
University of Twente, P.O Box 217, 7500 AE Enschede, The
Netherlands,\\
$^2$ Technical University of  Munich, Dept. of  Zoololgy, 
D-85747 Garching, Germany
}


\begin{abstract}
Fluid dynamics video:
Snapping shrimp produce a snapping sound by an extremely rapid closure of their snapper claw. Our high speed imaging of the claw closure has revealed that the sound is generated by the collapse of a cavitation bubble formed in a fast flowing water jet forced out from the claws during claw closure. The produced sound originates from the cavitation collapse of the bubble. At collapse 
a short flash of light is emitted, just as in single bubble 
sonoluminescence.
A model based on the Rayleigh-Plesset equation 
can quantitatively account for the visual and acoustical observations. 
\end{abstract}

\maketitle

The video
\href{http://ecommons.library.cornell.edu/bitstream/1813/9379/1/Gallery+Video.mpg}{``On the sound of snapping shrimp''} 
contains various sequences of the animal itself, the claw closure, 
the bubble formation and its collapse, and
an animatd explantion of the hydrodynamical model.

For further reading we refer to references \cite{ver00,loh01,ver01}.

\end{document}